\documentclass[prapplied, reprint,showpacs,  superscriptaddress]{revtex4-2}

\usepackage{graphicx}
\usepackage{dcolumn}
\usepackage{graphicx}
\usepackage{epsfig}
\usepackage{epsf}
\usepackage{amssymb,amsmath,amsthm,bm,bbm,dsfont}
\usepackage{empheq,cases}
\usepackage{mathtools}
\usepackage{multirow}
\usepackage[colorlinks=true,linkcolor=blue!50!black,citecolor=red,pdfauthor={ },pdftitle={ },pdfsubject={ },pdfkeywords={ }]{hyperref}
\usepackage{tikz}
\usepackage{pgfplots}
\usepackage{xcolor}
\usepackage{lipsum}

\usepgfplotslibrary{patchplots}
\usetikzlibrary{decorations.text }
\usetikzlibrary{shapes.symbols}
\usetikzlibrary{intersections,shapes.arrows}
\usetikzlibrary{decorations.markings}
\usetikzlibrary{mindmap,trees, backgrounds,fadings,shadows}
\usetikzlibrary{positioning,calc}
\usetikzlibrary{arrows,snakes,shapes}
\usetikzlibrary{patterns}

\definecolor{c1}{RGB}{219,68,56}
\definecolor{c2}{RGB}{74,91,163}
\definecolor{c3}{RGB}{105,182,106}

\theoremstyle{definition}

\usepackage{multirow}
\usepackage{dcolumn}
\newcolumntype{d}[1]{D{.}{.}{#1}}

\begin{document}

\title{Fast  Ion  Gates Outside the Lamb-Dicke Regime by Robust  Quantum Optimal Control }

\author{Xiaodong Yang}
\affiliation{Shenzhen Institute for Quantum Science and Engineering, Southern University of Science and Technology, Shenzhen, 518055, China}
\affiliation{International Quantum Academy, Shenzhen, China}
\affiliation{Guangdong Provincial Key Laboratory of Quantum Science and Engineering, Southern University of Science and Technology, Shenzhen, 518055, China}

\author{Yiheng Lin}  
\affiliation{CAS Key Laboratory of Microscale Magnetic Resonance and School of Physical Sciences, University of Science and Technology of China, Hefei 230026, China}
\affiliation{CAS Center for Excellence in Quantum Information and Quantum Physics, University of Science and Technology of China, Hefei 230026, China}
\affiliation{Hefei National Laboratory, University of Science and Technology of China, Hefei 230088, China}

\author{Yao Lu}
\email{luy7@sustech.edu.cn}
\affiliation{Shenzhen Institute for Quantum Science and Engineering, Southern University of Science and Technology, Shenzhen, 518055, China}
\affiliation{International Quantum Academy, Shenzhen, China}
\affiliation{Guangdong Provincial Key Laboratory of Quantum Science and Engineering, Southern University of Science and Technology, Shenzhen, 518055, China}

\author{Jun Li}
\email{lij3@sustech.edu.cn}
\affiliation{Shenzhen Institute for Quantum Science and Engineering, Southern University of Science and Technology, Shenzhen, 518055, China}
\affiliation{International Quantum Academy, Shenzhen, China}
\affiliation{Guangdong Provincial Key Laboratory of Quantum Science and Engineering, Southern University of Science and Technology, Shenzhen, 518055, China}

\begin{abstract}
We present   a robust quantum optimal control framework for implementing fast entangling gates on ion-trap quantum processors.  The framework   leverages   tailored laser pulses to    drive the multiple vibrational sidebands of the ions  to create phonon-mediated entangling  gates and, unlike the state of the art,  requires neither     weak-coupling Lamb-Dicke approximation nor   perturbation treatment. With the application of gradient-based
optimal control, it enables finding amplitude- and phase-modulated laser control protocols that work beyond the
Lamb-Dicke regime, promising gate speed at the order of microseconds comparable to the characteristic trap
frequencies. Also,   robustness requirements  on the temperature of the ions and initial optical phase can be conveniently included  to pursue  high-quality fast gates against experimental imperfections.
Our approach represents a step in   speeding up quantum
gates to achieve larger quantum circuits for quantum computation and simulation, and thus can find applications in near-future experiments. 
\end{abstract}

\maketitle

\section{Introduction}

Trapped atomic ions are considered a rather promising platform for realizing large-scale quantum computers  \cite{Blatt08,Monroe13,Sage19}. To bring this potential to reality, it is essential to perform fast, accurate, and robust entangling gates \cite{DiVincenzo2000}. While high-fidelity single-qubit gates  have been implemented and   two-qubit gates have also  steadily improved their precisions over the past decades \cite{PhysRevLett.117.060504,PhysRevLett.117.060505,PhysRevLett.125.150505,PhysRevLett.127.130505,srinivas2021},   the time-efficiency of performing entangling gates remains  exceedingly hard to scale faster. Current  implementations of entangling gates     typically cost   tens to hundreds of microseconds \cite{Wright2019,PRXQuantum.2.020343,PRXQuantum.3.010347}, much slower than the speed limit set by the trap period  of   a few microseconds. The rather slow gate speed hence represents a limiting factor for reaching larger-sized quantum      information processing experiments on ion-trap systems.

Entangling gates on trapped ions are realized through the mechanism of phonon-mediated  qubit couplings, which is created by laser driving of the vibrational modes of  the ions in the trap \cite{PhysRevLett.74.4091,PhysRevLett.82.1971,PhysRevLett.87.127901}. 
Most existing    gate   schemes  essentially rely on   the Lamb-Dicke approximation which assumes slow  field driving and weak ion-motion interactions, and also  on approximation technique of perturbation theory such as Magnus expansion.    When increasing the gate speed, much stronger laser driving is needed,  and the  resulting   larger displacements of the ions in phase space  cause considerable deviations from the Lamb-Dicke regime. Accordingly,  the   previous approximations are no longer valid that, the higher-order error terms in evolution that have not been   allowed for before now become   prominent. This poses the major hindrance in constructing fast gates outside the Lamb-Dicke regime.

There have been considerable theoretical \cite{PhysRevLett.91.157901,PhysRevLett.93.100502,steane2014pulsed,PhysRevA.101.052328,PhysRevA.103.052603,PhysRevResearch.3.013026,Kihwan22} and experimental \cite{PhysRevLett.119.230501,Lucas18,Blatt19}  efforts to speed up ion gates, the attempts   are challenging. For example, Ref. \cite{Lucas18} reported a high-fidelity (99.8\%) 1.6 $\mu$s gate and a much worse fidelity ($\sim$60\%) 480 ns gate in experiments. The  pulse synthesis method  adopted there    is  still under the Lamb-Dicke condition, so the breakdown of Lamb-Dicke approximation comprises a major source of gate infidelity. Theoretically, there were    proposals that  take more    Lamb-Dicke expansion terms into account \cite{PhysRevA.103.052603,Kihwan22},  but as    higher-order sideband transitions are driven, it gets more complicated to derive the corresponding suitable driving profiles.  Within the Lamb-Dicke regime, comparably
simple driving schemes can be devised; outside the Lamb-Dicke regime, the nonlinearity inherent in the  Hamiltonian   makes the dynamical control problem less tractable. Moreover, the theoretical analysis  can be even harder when one wants to add    consideration of robustness to various   noises. 

In this work, we propose to use robust  quantum optimal   control (QOC) to tackle the problem of fast ion gates. QOC is a flexible  and very effective method in finding high-performance pulses that accomplish given control tasks on a   quantum system. Over the many years, it   becomes a versatile tool in   quantum technologies \cite{Wilhelm22} and has found broad applications in diverse quantum platforms \cite{khaneja2005optimal,Wilhelm15,PhysRevLett.118.150503,PhysRevLett.103.110501,nv20,PhysRevLett.112.190502}. In particular,  QOC   plays an important role in  trapped-ions systems for devising and implementing  amplitude, frequency or phase-modulated entangling gates   \cite{PhysRevLett.114.120502,PhysRevLett.120.020501,Figgatt19,Lu19,Slatyer20},  but usually being used in combination with   Lamb-Dicke approximation and second-order Magnus expansion. Here, we show that, such approximations are unnecessary.   We      construct     a  general framework of QOC-based fast ion gates outside the Lamb-Dicke regime, which directly deals  with the full dynamical  evolution   generated by the full Hamiltonian.  The construction is exemplified      on   the M{\o}lmer-S{\o}rensen scheme \cite{PhysRevLett.82.1971}, but should apply well to other similar gate schemes.  Robustness  requirements are also   incorporated into   QOC   as either extra constraints on pulse parameters or additional optimization objectives.  We then give a concrete  two-qubit gate example   with duration 3 $\mu$s, infidelity $< 10^{-3}$ and insensitivity to initial optical phase,      which demonstrates the applicability of  our framework.
  We see that an advantage of fast gates is that they are less sensitive to errors associated with motional frequency dirfts or heating due to their   significantly reduced gate   time. 

\section{Problem Description}
\begin{figure}[t]
\includegraphics[width=\linewidth]{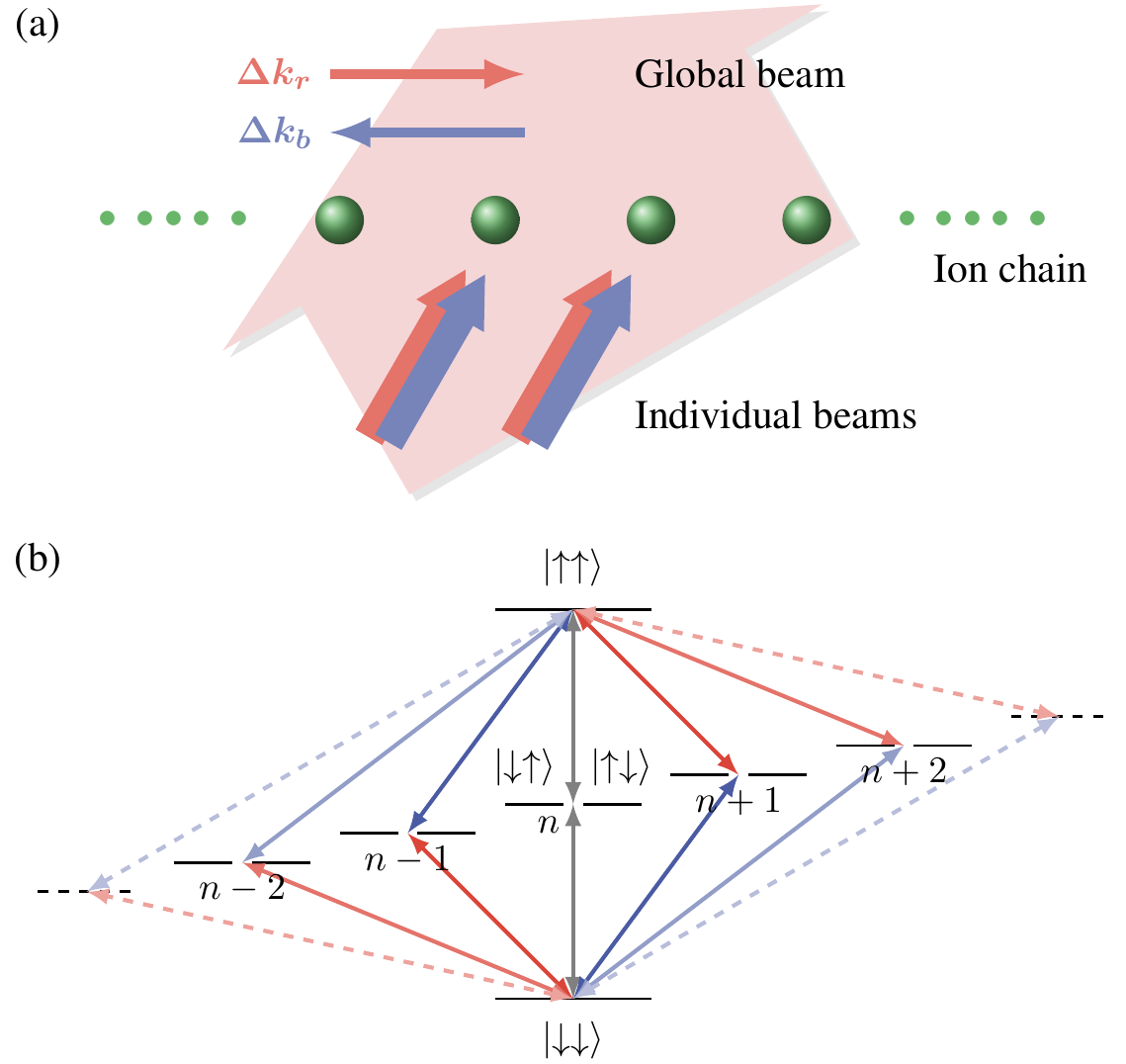}
\caption{  (a) Laser beam geometry in phase insensitive configuration. The wavevector differences for pairs of frequencies driving the red sideband and the blue sideband travel in the opposite direction. (b) Illustration of multi-sideband driving protocol under the M{\o}lmer-S{\o}rensen gate scheme. Qubits are stored in the hyperfine states of the ions denoted by $\left|\uparrow\right\rangle$ and $\left|\uparrow\right\rangle$, and the   motional state is characterized by the phonon number $n$. Outside the Lamb-Dicke regime, higher-order blue and red sideband transitions are also excited simultaneously.}
\label{fig-1}
\end{figure}

Consider a single chain consisting of $N$ identical ions which are one-dimensionally aligned along $z$ axis; see Fig. \ref{fig-1}(a). Qubits are encoded in a pair of internal energy levels belonging to each ion, while the whole ion chain vibrates collectively due to the long-range Coulomb repulsion. Therefore, the free Hamiltonian of the above system is   ($\hbar = 1$) 
\begin{equation}
   \hat H_0 = 
        \sum_{k = 1}^{N} \omega_q \hat\sigma^k_{z}/2 + 
        \sum_{j = 1}^{N} \nu_{j} \hat a_{j}^\dag \hat a_{j}.
\end{equation}
Here, $\omega_{q}$ denotes the energy gap of the encoded ion qubits, and $\hat \sigma_{x,y,z}^{k}$ are Pauli matrices for the $k$th qubit. For simplicity, we shall only consider the collective motional modes along the $z$ axis, and $\nu_{j}$ is the eigen-frequency of the $j$th mode with $\hat a_{j}^\dag$ and $\hat a_{j}$ being the corresponding ladder operators.    
The ion qubits are conventionally coupled to the motional modes via laser-ion interactions. Individual addressing ability on each qubit is assumed in our general discussion, as illustrated in Fig. \ref{fig-1}(a). Although this diagram considers hyperfine qubits manipulated via the two-photon stimulated Raman transitions, our approach is suitable for optical qubits as well.

We now describe the widely used M{\o}lmer-S{\o}rensen \cite{PhysRevLett.82.1971,PhysRevLett.82.1835} gate  scheme. In its basic form, it requires bichromatic laser fields with frequencies of $\omega_q \pm \omega$ to illuminate targeted ions and simultaneously induce the first blue and red sideband transitions. We here assume the phase-insensitive configuration of the bichromatic fields, where the effective wave vectors for the two sidebands ($\Delta \bm{k}_b = -\Delta \bm{k}_r = \bm{k}$) propagate in opposite directions \cite{Monroe05, Kihwan22}.  As a result, we can   obtain  the  following expression for the    laser-ion interaction Hamiltonian   in the ions'  resonant rotating frame
\begin{equation}
\hat H_C(t)    = \sum_{k=1}^N \dfrac{\Omega}{2} \left [
            e^{ i \left (
                -\bm{k}  z_{k} - \omega t + \phi_{r}
            \right )} + 
            e^{i\left (
                \bm{k}  z_{k} + \omega t + \phi_{b}
            \right ) }
        \right ]
       \hat  \sigma_{+}^k + \text{h.c.}   \nonumber 
\end{equation}
Here, the effective laser-ion coupling strengths for the both sideband transitions are balanced to an equal value of $\Omega$, while $\phi_b$ and $\phi_r$ represent the independent optical phase for the blue and red sideband, respectively. 
If we further define the so-named spin phase $\phi = (\phi_{b} + \phi_{r})/2$ and the motional phase $\varphi = (\phi_{b} - \phi_{r})/2$ \cite{Monroe05}, the above equation  can be rewritten as 
\begin{equation}
  \hat H_C(t) = \sum_{k=1}^N \Omega
        \hat \sigma^k_{\phi_S} \cos{ \left (
            \omega t + \varphi + \bm{k} z_{k} 
        \right )}, \nonumber
\end{equation}
where $\hat \sigma_{\phi} = \hat \sigma_{x} \cos\phi + \hat \sigma_{y}\sin\phi$. Note that, the pulse parameters of $\{ \Omega, \phi, \varphi \}$ can be site-dependent if one of the beams to drive the Raman transitions has the ability of   individual addressing. The position operator $z_k$ can be  expanded into the linear combination of the operators of the collective motion as $  z_k = \sum_{j} b_{kj} \Delta  z_j ( \hat a_j + \hat a_j^\dagger)$, where $\Delta   z_j = \sqrt{\hbar/2M\nu_j}$ is the size of the ground-state wavepacket of the $j$th motional mode, and ${b_{kj}}$ is the normal-mode transformation matrix. Conventionally, we denote      the coefficient $b_{kj}\bm{k}\Delta z_{j}$ by $\eta_{kj}$, which are referred to as the Lamb-Dicke parameters. Accordingly, the control Hamiltonian can be further expressed as 
\begin{equation}
 \hat    H_C(t) = \sum_{k=1}^N \Omega  \hat \sigma_{\phi}^{k}
        \cos{\left[
            \omega t + \varphi + \sum_{j=1}^N \eta_{kj}( \hat a_j +  \hat a_j^\dagger)  
        \right]}
        . \nonumber
\end{equation}
If the ions are sufficiently cooled and the spatial motion keeps small, so that the condition for the Lamb-Dicke regime is met, then the laser-ion interaction Hamiltonian   admits  a simplified form by taking the approximation $e^{i\eta (\hat a + \hat a^\dag)}\approx 1 + i \eta (\hat a + \hat a ^\dag)$. This is the normal way to    simplify treatments of   laser driven    ion dynamics. Most of the currently employed entangling gates operate   in   the Lamb-Dicke regime.

To go beyond the Lamb-Dicke regime, it is desirable to    excite the higher-order sideband transitions. Hence we would like to  use  multichromatic   laser beam    to     address the multiple sidebands simultaneously; see an illustration in Fig. \ref{fig-1}(b). Such a driving protocol was   previously  used to generate robust gates against different sources of noises \cite{PhysRevLett.121.180502,PhysRevA.101.032330}. For the present problem, we assume that the multichromatic   laser beam   contains $L$ frequency pairs  $ \{\pm \omega_l: l=1,...,L \}$, and for the $l$th frequency component, its corresponding amplitude, spin phase  and motional phase modulations are denoted by $\Omega_l(t)$, $\phi_l(t)$ and $\varphi_l(t)$, respectively. Similar to the previous derivations, the   control Hamiltonian now has the form
\[
	\hat H_C  =      \sum_{k=1}^N   \sum_{l=1}^L  \Omega_l \hat \sigma_{\phi_l}^{k} \cos\left[\omega_l t  + \varphi_l + \sum_{j=1}^N \eta_{kj}(\hat a_j + \hat a_j^\dagger) \right].  
\]
We point out that, although we   assume predefined driving frequencies in this work, our method can apply as well to continuous frequency modulation  in which the pulse frequency serves as optimization parameters.

Our goal is to   perform  a target quantum gate   $\hat{\overline U} $, which is  a unitary transformation on the computational states namely the internal states. 
The   task is hence to find a shaped control pulse determined by a set of real-valued functions $\{\Omega_l(t), \phi_l(t), \varphi_l(t)\}$ such that, the time propagator  at the end of the evolution   implements   $\hat{\overline U} $ on the ions' internal state, but meanwhile should not  change the ions' external state. If the ions can be perfectly cooled, we can assume the ions are initially at the vibrational ground state. But this is a difficult condition. More realistically, the motional state is at   a statistical mixture. In this work, we shall suppose that the ions' vibrational modes are initially in the thermal product state $\hat \rho_\text{th} =   \hat \rho^1_\text{th} \otimes \hat  \rho^2_\text{th} \otimes \cdots \otimes \hat \rho^N_\text{th}$ with   
\[\hat  \rho^j_\text{th} = \left(1 - e^{-\nu_j/k_B T} \right) \sum_{n\ge 0}  |n\rangle\langle n| e^{- n \nu_j/k_B T}    \]
for $j=1,...,N$, where $k_B$ is the Boltzmann constant and $T$ represents the temperature. Recall that the Lamb-Dicke regime refers to the situation where the spatial spread of the wave-function of the ions is much smaller than the wavelength of the laser field, or in mathematical words, $\eta \sqrt{\bar n +1} \ll 1$ where $\bar n$ is the average phonon occupation number.  But  to implement  fast gates,   strong laser driving is necessary,  and   the ion motion may be considerably excited during the evolution, so the   Lamb-Dicke approximation would no longer  be valid. Besides, to avoid imposing stringent requirements on ion cooling, we would like   our   optimal control scheme be      effective    without necessarily 
assuming  negligible initial average occupation $\bar n$.

%
%
%
%
%

\section{Quantum Optimal Control}
In this section, we shall formulate the quantum optimal control problem for the  task of  fast ion gate outside the Lamb-Dicke regime, and describe how to solve the problem by using   gradient-based methods.

\subsection{System and Control}
As the first step, we  make clear the   system Hamiltonian and the control Hamiltonian. 
In the ions' resonant rotating frame,     the system Hamiltonian goes    
\begin{equation}
	\hat H_S   = \sum_{j=1}^N \nu_j \hat a_j^\dag \hat a_j, \label{H_S}
\end{equation}
and the control Hamiltonian    can be written as
\begin{align}
	\hat H_C & =      \sum_{k=1}^N   \sum_{l=1}^L  \Omega_l \hat \sigma_{\phi_l}^{k} \cos\left[\omega_l t  + \varphi_l + \sum_{j=1}^N \eta_{kj}(\hat a_j + \hat a_j^\dagger) \right]  \nonumber \\
	& =\sum_{k=1}^N   \sum_{l=1}^L  \Omega_l \hat \sigma_{\phi_l}^{k} \left[\cos(\omega_l t  + \varphi_l)  \hat \xi^k_x + \sin(\omega_l t  + \varphi_l)  \hat \xi^k_y \right], \nonumber 
\end{align} 
where 
\begin{align}
\hat \xi^k_1 & = \cos\left[ \sum_{j=1}^N \eta_{kj}(\hat a_j + \hat a_j^\dagger) \right], \nonumber \\
\hat \xi^k_2 & = -\sin\left[ \sum_{j=1}^N \eta_{kj}(\hat a_j + \hat a_j^\dagger) \right]. \nonumber	
\end{align}
To   simplify   the notations in our subsequent derivations, we define  $\hat \xi^k_{\omega_l t  + \varphi_l} = \cos(\omega_l t  + \varphi_l)  \hat \xi^k_1 + \sin(\omega_l t  + \varphi_l)  \hat \xi^k_2$, then
\begin{equation}
\hat H_C 	= \sum_{k=1}^N   \sum_{l=1}^L  \Omega_l \hat\sigma_{\phi_l}^{k} \hat\xi^k_{\omega_l t  + \varphi_l}. \label{H_C}
\end{equation}
Assume that the laser pulse has   time length $T$, then the combined evolution of $H_S + H_C(t)$   results in the   time evolution operator 
\begin{equation}
	\hat U(T) = \mathcal{T}\exp \left(-i \int_0^T dt [\hat H_S + \hat H_C(t)  ] \right),
\end{equation}
where $\mathcal{T}$ denotes the time ordering operation.
Since  our approach no longer takes  the Lamb-Dicke approximation of any order, all  the nonlinear dependence
of $\hat H_C$ on the phonon creation and annihilation operators will contribute to the controlled evolution.


\subsection{Average Gate Fidelity}
In quantum optimal control, normally we define an objective function to measure  the control performance. Note that, while   $\hat U(t)$ is unitary in the whole state space, it   generates a quantum channel in the subspace of internal states. How well this channel realizes the target gate can be quantitatively characterized through  the average gate fidelity function,   which we describe as follows. Suppose the  initial input internal state of the ions is $|\psi\rangle\langle \psi| $,   suppose  the ions are initially at the thermal state $\rho_\text{th}$, then at the end of the controlled evolution the actual output internal state is given by $\hat \rho = \operatorname{Tr}_\text{m}\left[ \hat U(T) |\psi\rangle\langle \psi| \otimes \hat \rho_\text{th} \hat U^\dag (T) \right]$, where $\operatorname{Tr}_\text{m}$ means taking partial trace over the external degrees of freedom.  The similarity between  $\hat \rho$ and the ideal state $\hat {\overline{U}} |\psi\rangle$ can be estimated by the state fidelity $\langle \psi| \hat {\overline{U}}^\dag \hat\rho \overline U |\psi\rangle $.
To compare $\hat U(T)$ and $\hat {\overline U}$, we need a state-independent measure which serves as our fitness function, that is, the average gate fidelity, defined by
\begin{equation}
	f = \int d\psi \langle \psi | \hat {\overline{U}}^\dag \operatorname{Tr}_\text{m} [ \hat U(T) |\psi\rangle\langle \psi| \otimes \hat \rho_\text{th} \hat U^\dag(T)  ] \hat {\overline{U}} |\psi\rangle, \nonumber
\end{equation}
where the integration is taken over the   uniform  (Haar)  measure $d\psi$ on the internal state   space.  A simple   expression for computing the average gate fidelity is the following \cite{Nielsen02}
\begin{equation}
	f = \frac{\sum_i \operatorname{Tr}\left\{ \hat {\overline{U}} \hat P_i^\dag \hat {\overline{U}}^\dag \operatorname{Tr}_\text{m}\left[ \hat U(T) \hat P_i \otimes \hat \rho_\text{th} \hat U^\dag (T) \right] \right\} + d^2}{d^2(d+1)}, \nonumber
\end{equation}
where $d=2^N$ and $\{\hat P_i\}$ is an orthogonal basis of $d \times d$ unitary operators such that $\operatorname{Tr}[\hat P_i^\dag \hat P_j] = \delta_{ij}d$. Here, we would   choose the Pauli   operators $\{\hat P_i\} = \{I, \hat \sigma_x,\hat \sigma_y,\hat \sigma_z \}^{\otimes N}$, where $I$ is the identity operator.

At this point, we are ready to formally state the quantum optimal control problem for target    gate realization: we set a suitable choice of the laser driving frequencies $\{\omega_l\}$, and
\begin{align}
\text{find}  \quad & \{\Omega_l(t), \phi_l(t), \varphi_l(t)  \}, \nonumber \\
	\max  \quad  & f(\hat U(T), \hat {\overline{U}}), \ \nonumber \\
	\text{s.t.} \quad   &  \frac{\partial \hat U(t)}{\partial t} = -i [\hat H_S + \hat H_C(t) ] \hat U(t). \nonumber
\end{align}


\subsection{Gradient-based Optimization}
In general, it is difficult to   construct   analytic optimal control solutions, so taking    numerical approach   is necessary. The GRAPE algorithm \cite{khaneja2005optimal} is   one     widely used    numerical optimization technique  for solving optimal control problems. In the following, we outline its basics.

First,  we should discretize the controlled     evolution. Let the evolution    be divided into $M$ slices of equal length $\tau = T/M$, and  in each time slice the control is time-invariant.
The time discretization $\tau$ is naturally given
by the time resolution of the lasers generating the pulse. 
We represent the  set of pulse parameters after discretization by an array $\bm{u} = (\bm{\Omega},\bm{\phi},\bm{\varphi})$, where $\bm{\Omega} = (\Omega_{lm})$, $\bm{\phi} = (\phi_{lm})$  and $\bm{\varphi} = (\varphi_{lm})$ are   arrays of pulse parameters all of size $L\times M$, with their $lm$th elements corresponding to the amplitude, spin phase and motional phase of the $l$th frequency component of the   pulse at the $m$th time slice,  respectively. Provided that $\tau$ is small, we can view the control Hamiltonian as constant in each slice. Let $\hat H_C[m]$ denote   the control Hamiltonian at the $m$th slice,  and $\hat U_m = \exp\left\{-i  \tau  (\hat H_S + \hat H_C[m] ) \right\} $ the corresponding   time evolution operator, then the total time evolution operator is   given by $\hat U(T) = \hat U_M \cdots \hat U_1$. 

To seek  an optimal pulse solution,  the GRAPE algorithm starts from an initial pulse guess $\bm{u}^{(0)}$, and generates a sequence of iterates  $\bm{u}^{(1)}, \bm{u}^{(2)}, ...,$ by taking steps along the gradient ascent direction
\begin{equation}
	\bm{u}^{(k+1)} = \bm{u}^{(k)} + \alpha^{(k)} \bm{g}^{(k)},
\end{equation}
where $\bm{g}^{(k)}$ is the gradient of the target function $f$ at the $k$th iterate $\bm{u}^{(k)}$, and  $\alpha^{(k)}$ is a   step size chosen such that an adequate increase in $f$ along $\bm{g}^{(k)}$ can be acquired. 
The gradient  $\bm{g}$ of $f$ with respect  to the control parameters can be evaluated according to
\begin{equation}
\bm{g}	=   \frac{\sum\limits_i \operatorname{Tr}\left\{ \hat {\overline U} \hat P_i^\dag \hat {\overline U}^\dag \operatorname{Tr}_\text{m}\left[\displaystyle \frac{\partial \hat U}{\partial \bm{u} } \hat P_i \otimes \hat \rho_\text{th} \hat U^\dag + \text{h.c.} \right] \right\}}{d^2(d+1)}, \label{gradient-1}
\end{equation}
where $\partial \hat U/\partial \bm{u} = (\partial \hat U/\partial \bm{\Omega}, \partial \hat U/\partial \bm{\phi}, \partial \hat U/\partial \bm{\varphi})$. Assuming that the  discretization step $\tau$ is small, and let $u[m]$ denote any of the control parameters    at the $m$th slice evolution, there is
\begin{align}
	\frac{\partial \hat U}{\partial u[m]} & =\hat  U_M \cdots \frac{\partial \hat U_m}{\partial u[m]} \cdots \hat U_1  \nonumber \\
	& = \hat  U_M \cdots \left(-i\tau \frac{\partial \hat H_C[m]}{\partial u[m]} \hat U_m \right) \cdots \hat U_1 + O(\tau^2).   \label{gradient-2}
\end{align}
Then, from Eq. (\ref{H_C})  we   have that
\begin{subequations}
\label{gradient-3}	
\begin{align}
\frac{\partial \hat H_C[m]}{\partial \Omega_{lm}}  & =	\sum_{k=1}^N        \hat \sigma_{\phi_{lm}}^{k} \hat \xi^k_{\omega_l m\tau  + \varphi_{lm}}, \\
\frac{\partial \hat H_C[m]}{\partial \phi_{lm}}  & =	\sum_{k=1}^N      \Omega_{lm}\hat \sigma_{\phi_{lm}+\pi/2}^{k} \hat \xi^k_{\omega_l m\tau  + \varphi_{lm}}, \\
\frac{\partial \hat H_C[m]}{\partial \varphi_{lm}}  & = \sum_{k=1}^N      \Omega_{lm}\hat \sigma_{\phi_{lm}}^{k} \hat \xi^k_{\omega_l m\tau  + \varphi_{lm}+\pi/2}. 
\end{align}
\end{subequations}
Hence, Eqs. (\ref{gradient-1}), (\ref{gradient-2}) and (\ref{gradient-3}) together provides an explicit way of  gradient evaluation.

In the practice of GRAPE, some important considerations need be made clear as below.

\emph{Smoothness consideration.} Real waveform generators have
finite response times. To avoid sharp edges as is the problem for    rectangular pulses,  we could restrict to consider the  following class of pulses using truncated  Fourier basis  
\begin{equation}
	\Omega(t)  = \sum_{k=1}^{k_{\max}} c_k \left[1 - \cos (2\pi k t/T)\right].
\end{equation}
Apparently,   pulse waveforms thus specified automatically goes smoothly to zero at the beginning and end of the pulse. Another simple strategy is to use a lowpass   filter to suppress  the high-frequency components in the pulse waveform after each iteration and also, the gradient needs  be   filtered as well.

\emph{Accuracy in gradient estimation}. For sufficiently small $\tau$, it suffices to take the first-order approximation in the expression Eq. (\ref{gradient-2}) for computing $\bm{g}$. It is also possible to    allow for the higher order terms to get more accurate estimation of $\bm{g}$ \cite{Kuprov11}, which   we will not brief here.

\emph{Convergence speed}.  Since GRAPE utilizes only first-order gradient information, it actually has   a linear convergence speed. Faster convergence  can be achieved if one also takes into account of higher-order gradients. For example,  conjugated
gradient  or quasi-Newton methods combine first- and second-order gradient information to determine a suitable gradient-related search direction, and thus can provide superlinear convergence speed.  Such   methods have   been introduced and practiced in quantum optimal control, resulting in  some improved variants of GRAPE \cite{BSV08,Kuprov11}. Interested readers are    referred to standard books on numerical optimization such as Ref. \cite{NW06} for more details.


\subsection{Initial Phase Robustness}
The above optimization presumes that we can control and stabilize the laser   phases, which is however not that true in real experiments. Specifically, in the phase-insensitive configuration considered here, there would exist an unpredictable initial motional phase $\varphi_0 ]$. This  means that the actual evolution is driven by the control Hamiltonian $\hat H_C(\Omega, \phi, \varphi + \varphi_0)$ rather than  $\hat H_C(\Omega, \phi, \varphi)$. To address  the  issue, it is   required that  the optimal pulse should  be  robust to $\varphi_0$ over the entire range $[0,2\pi]$.

Our consideration of initial phase robustness consists of two steps. First, we sample  a set of different initial optical phases $\{ \varphi^s_0\in [0,2\pi]\}$   and compute their corresponding gate fidelities $\{f_{\varphi^s_0}\}$,   here $s=1,..,S$ with $S$   the number of samples. It is   natural to use the    average of the gate fidelities over the sampled initial phases  $\frac{1}{S}\sum_{s=1}^S f_{\varphi^s_0}$    as the new objective function. Next, we further require that the controlled evolution has first-order robustness property with respect to $\varphi_0$  at the points of the sampled phases.

We remark that both steps are important. If we only use sampling, the obvious drawback is that  we can only make the fidelity optimal at   the sampled phases, while for the other unsampled phases the fidelity could very likely be still low. On the other hand, if we only employ robust control, it is necessary to involve complicated higher-order perturbation theory to ensure that the robustness region can cover the entire region $[0,2\pi]$. Therefore, we are inclined to   combining  both techniques to achieve   smooth, wide-range pulse robustness.

The sampling-based optimization step is quite straightforward. Now, we describe how to incorporate first-order robustness into optimal control. As illustration, we consider $\varphi_0$ robustness at $\varphi_0=0$. Suppose there is a small perturbation $\delta\varphi$, the control Hamiltonian becomes
\begin{align}
\hat H_C &	= \sum_{k=1}^N   \sum_{l=1}^L  \Omega_l \hat\sigma_{\phi_l}^{k} \hat\xi^k_{\omega_l t  + \varphi_l +\delta \varphi} \nonumber\\
 & = \sum_{k=1}^N   \sum_{l=1}^L  \Omega_l \hat\sigma_{\phi_l}^{k} (\hat\xi^k_{\omega_l t  + \varphi_l} + \delta \varphi \hat\xi^k_{\omega_l t + \varphi_l + \pi/2} + O(\delta\varphi^2)). \nonumber
\end{align}
Hence, to first-order approximation   the perturbed Hamiltonian is
\begin{equation}
	\hat H_S+\hat H_C(t) +\delta \varphi \hat H'(t), \nonumber
\end{equation}
where we define 
\[\hat H'(t) = \sum_{k=1}^N   \sum_{l=1}^L  \Omega_l \hat\sigma_{\phi_l}^{k} \hat\xi^k_{\omega_l t  + \varphi_l +\pi/2}\] as the perturbation operator. Without perturbation, the system evolution is given by $\hat U(t)$ as before; in the presence of perturbation, the real evolution $\hat U_\text{real}(t)$ deviates from the ideal $\hat U(t)$, and the deviation can be characterized by the Dyson series \cite{PhysRev.75.486} $\hat U_\text{real}(t) = \hat U(t) + \hat U_\text{err}^{(1)}(t) +    \cdots$. Here, we shall only keep the first-order error term $U_\text{err}^{(1)}(t)$, which  has the expression
\begin{equation}
\hat U_\text{err}^{(1)}(t) = -i\hat U(t)\int_0^t	 dt_1 \hat U^\dag(t_1) \delta \varphi \hat H'(t_1) \hat U(t_1). \nonumber
\end{equation}
One can introduce the  definition
\begin{align}
\mathcal{D}_{\hat U(t)}(\hat H'(t)) & = \frac{\hat U_\text{real}(t) - \hat U(t)}{\delta\varphi} \nonumber \\
&= -i\hat U(t)\int_0^t	 dt_1 \hat U^\dag(t_1)   \hat H'(t_1) \hat U(t_1).	\nonumber
\end{align}  
It  captures the first-order perturbative effect  in the evolution  operator  $\hat U(t)$ due to the presence of $\hat H'(t)$, and is  referred to as the  directional derivative of $\hat U(t)$ along  $\hat H'(t)$ in Ref. \cite{HPZC19}.  Therefore, minimizing the magnitude of $\mathcal{D}_{\hat U(t)}(\hat H'(t))$ would imply that even there is a small phase variation $\delta \varphi$, the actual evolution is still close to the ideal one. Ref. \cite{HPZC19} also provides a very effective way to compute $\mathcal{D}_{\hat U(t)}(\hat H'(t))$ and its gradient, based on the technique of Van Loan matrix integral.

\section{Two-qubit Example} 
Now, we give an explicit two-qubit gate example to show the effectiveness of our method. We consider two $^{171}$Yb$^{+}$ ions in a trap frequency of $\nu_z = 2\pi \times 1$MHz along $z$-axis \cite{PhysRevA.76.052314}; therefore, the motional frequencies of the ion-chain turn out to be $\nu_1 = \nu_z$ and $\nu_2 = \sqrt{3}\nu_z$ for the center-of-mass mode and stretching mode, respectively. The ion qubit is encoded in the hyperfine clock state in the ground manifold and then manipulated via 355~nm lasers \cite{PhysRevLett.112.190502,PhysRevLett.120.020501,PhysRevApplied.13.024022}, giving the Lamb-Dicke parameter around $0.136$ for the center-of-mass mode.


The target gate that we consider in our numerical tests is      the maximally entangling gate $  \hat U_{XX}   = \exp\left(i \pi \sigma_1^x \sigma_2^x/4 \right)$ for the internal degrees of freedom.    Our driving protocol can be targeted at   the sidebands of  either the center-of-mass mode or the stretching mode, though the other mode would also be partially excited as we are in the fast gate regime. For simplicity, we shall consider amplitude-modulation only, with all  spin phases and    motional phases   set to   zero.

\begin{figure}[t]
\includegraphics[width=0.95\linewidth]{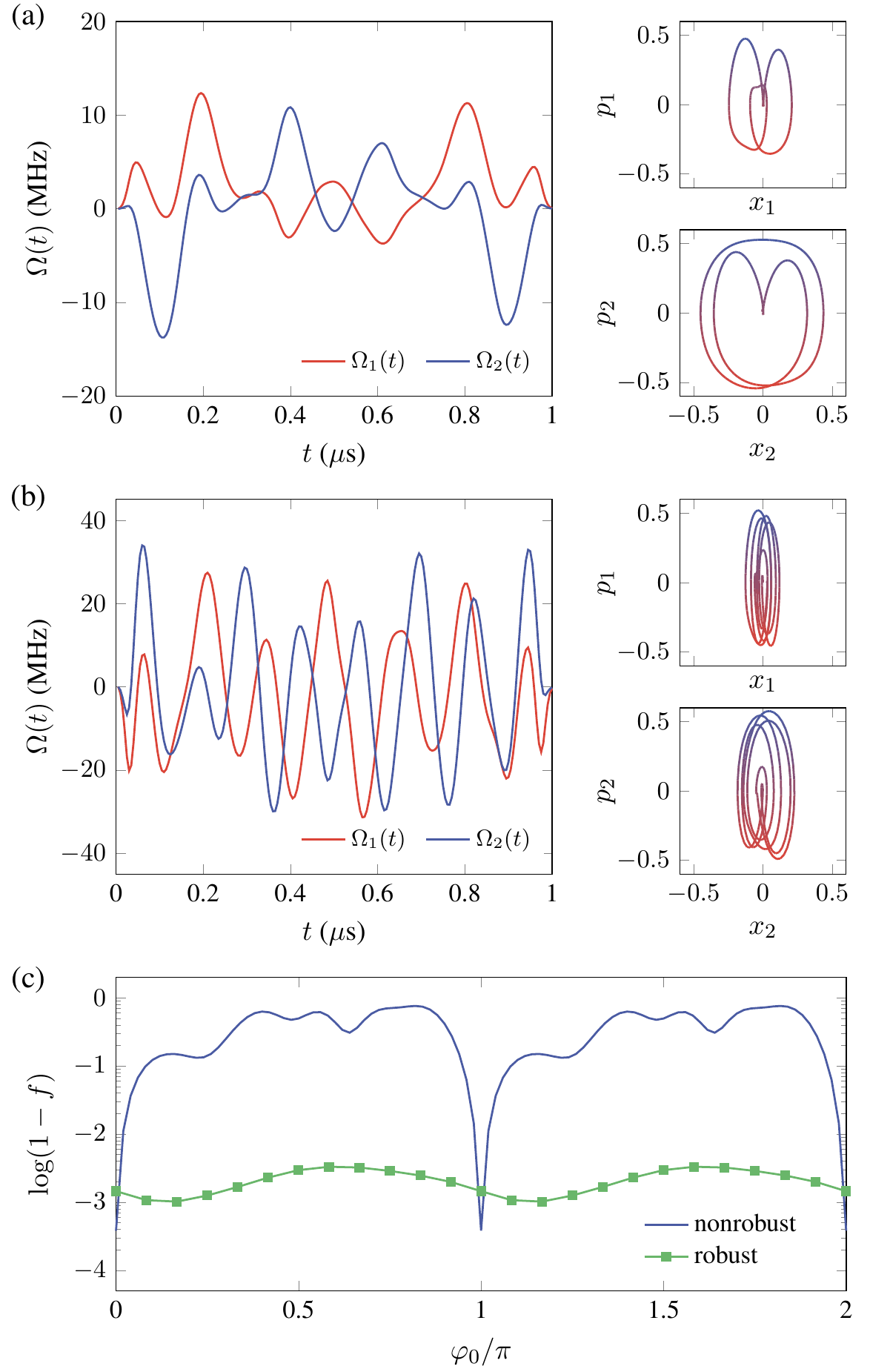}
\caption{Optimal laser control pulses   for the entangling gate $\hat U_{XX}$ with gate time $T=1$ $\mu$s and time step $\tau=5$ ns, under the ideal condition of perfect cooling $\bar n_1= \bar n_2 =0$. (a) An optimal pulse solution of  fidelity 0.9996, under the    assumption that all phases are stabilized at zero. (b) A robust optimal pulse that is insensitive to initial phase instability, having   phase-averaged gate  fidelity $\sim 0.9979$. All the  time-dependent amplitude shapes     start  and end  smoothly at zero. Negative amplitudes  correspond  to    an   inversion   of their  corresponding spin phases.  (c) Performance comparison of the two pulses in (a) and (b) in terms of initial phase robustness.}
\label{result_1us}
\end{figure} 

\begin{figure*}
\includegraphics[width=0.95\linewidth]{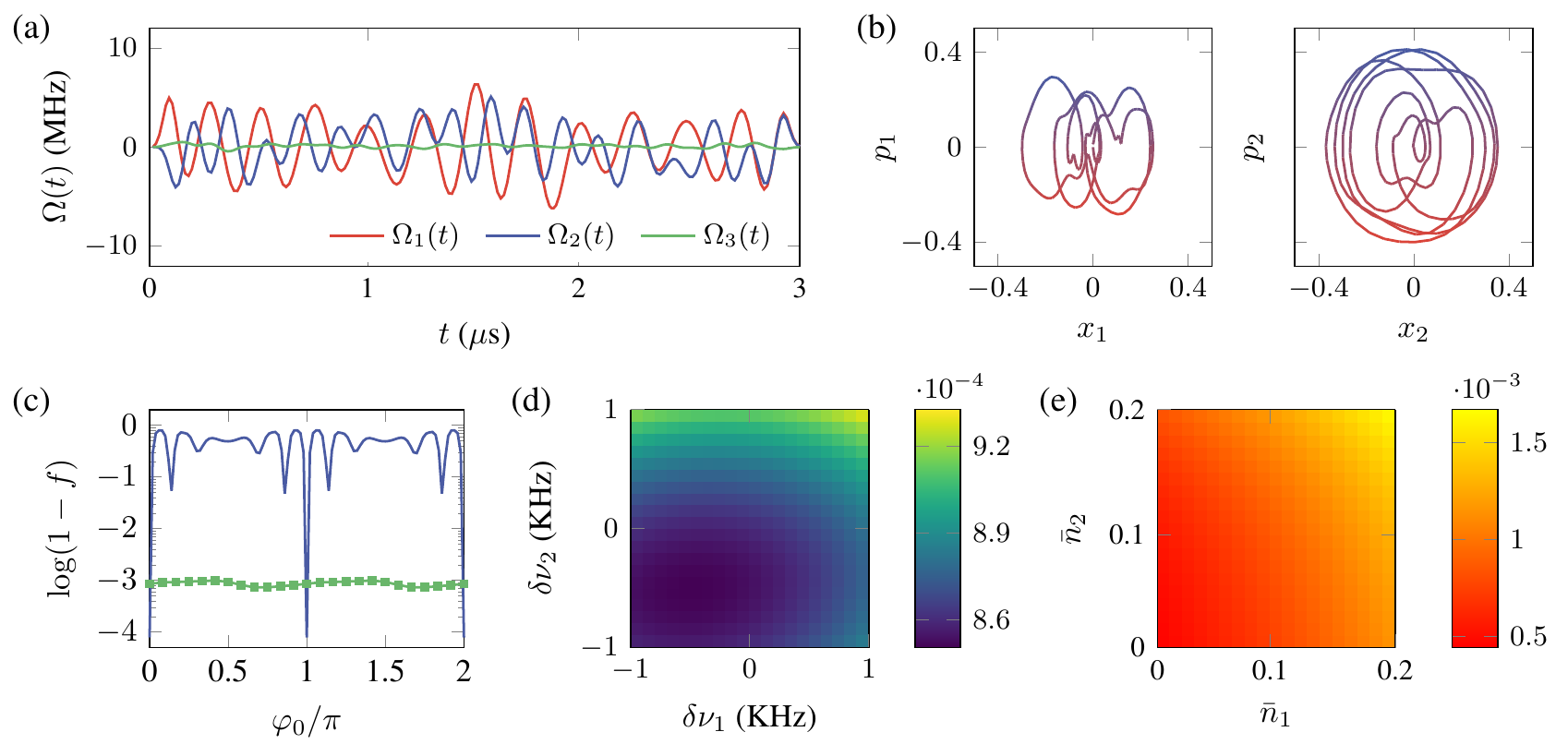}
\caption{Optimal laser control pulse   for the entangling gate $\hat U_{XX}$ with gate time $T=3$ $\mu$s and time step $\tau=15$ ns, under the condition of initial thermal occupation $\bar n_1= \bar n_2 =0.1$. 
(a) Pulse shapes. The $l$th time-varying shape $\Omega_l(t)$ represents the     amplitude modulation corresponding to the frequency component  $\omega_l =   l \nu_2$ for  $l=1,2,3$. 
(b) Phase-space trajectories. 
These are overlapping at zero, indicating no residual state-motional entanglement. (c) Initial phase robustness.    (d) Robustness profile with respect to  normal-mode frequency errors. (e) Robustness profile with respect to initial thermal occupation.}
\label{result_3us}
\end{figure*}

\subsection{1 $\mu$s-gate}
We  study the  case of gate time  $T=1$ $\mu$s at the start,  which corresponds    to one vibrational  period of the   center of mass mode  $ 1/ \nu_1$. We first assume pulse control under ideal   conditions      that the   ions' motion is   cooled perfectly,   the pulse amplitudes can be arbitrarily large,   and    pulse  phases can be made stable.  
Figure \ref{result_1us}(a) shows  a typical result  of   our   obtained optimal   pulse   for implementing $\hat U_{XX}$, which achieves an average gate fidelity of 0.9996. It has two pairs of frequencies $\{\pm \omega_l: \omega_l =   l \nu_1; l=1,2 \}$, as we find that using a single pair of frequencies is not sufficient to get a high fidelity. This is reasonable, since in the fast gate regime   the  Lamb-Dicke approximation is no longer valid, so it is needed to excite the higher-order sidebands.  
To show that our pulse    indeed does not  induce residual qubit-motion entanglement at the conclusion of the control operation, we plot the phase-space displacements of the two motional modes during the operation; see  Fig. \ref{result_1us}(b). 
The phase-space trajectories are calculated via 
\begin{align}
	x_j(t) &   = \operatorname{Tr} \left( {\hat \rho(t)      \left| \psi_j \right\rangle \left\langle \psi_j  \right| \otimes \hat x_j } \right), \nonumber \\
	p_j(t) &  = \operatorname{Tr} \left( {\hat \rho(t)      \left|\psi_j  \right\rangle \left\langle \psi_j  \right| \otimes \hat p_j } \right), \nonumber
\end{align}
where $|\psi_j\rangle$ is $\left|++ \right\rangle$ for mode $j=1$ or   $ \left|+- \right\rangle$ for mode $j=2$, $\hat x = (\hat a+ \hat a^\dag)$ and $\hat p = i(\hat a - \hat a^\dag)$,  $\hat \rho(t)$ is the state of the total system at time $t$ with the initial state being   $\left|\uparrow \uparrow\right\rangle \left\langle \uparrow \uparrow \right| \otimes \hat \rho_\text{th}$.   

Next,    we consider robustness to initial phase $\varphi_0$ in   our pulse search procedure. We choose four samples $\varphi_0 \in \{0,\pi/4,\pi/2,3\pi/4\}$ and perform optimization such that the   gate fidelity averaged over these samples is as high as possible.    An optimal pulse solution is given in Fig. \ref{result_1us} (b), which achieves fidelity $\sim 0.9979$. It has   larger amplitudes, and the resulting phase space trajectories are   more complex. Fig. \ref{result_1us}(c) compares initial phase robustness of the robust pulse and the  previous one. We can see that   robustness has indeed been  substantially improved, but at the cost of some infidelity increase.

Our numerically found two optimal pulse solutions     have maximum strength   reaching a few tens of MHz.  In practice,   however, such pulses may  be not experimentally friendly.  It seems hard to find better performance 1 $\mu$s pulses, hence to make further improvements,    we need to increase the pulse length.

\subsection{3 $\mu$s-gate}
In order to lower the overall laser intensity, we set a longer pulse time $T=3$ $\mu$s and perform   pulse search under realistic experimental conditions. 
Concretely, we   assume that the initial  thermal state $\hat \rho_\text{th}$ for the two motional modes has average occupation  of $\bar n_1 = 0.1$ and $\bar n_2 = 0.1$,    which is well within experimental capability. Robustness to initial phase $\varphi_0$
is also required. Figure \ref{result_3us} shows  a   result  of   our   obtained robust optimal   pulse, which achieves an average gate fidelity of 0.9991.  Figure \ref{result_3us}(c) compares       initial phase robustness  performances between pulses  before and after robustness optimization.  We also evaluate the robustness of  the gate     with respect to drifts in the two motional mode frequencies, finding that the pulse can suppress errors
in gate fidelities to below $1 \times 10^{-4}$ for up to a $\pm 1$ KHz frequency offset.   
In Fig. \ref{result_3us}(e), we plot   the gate infidelity   with respect to the initial thermal occupation $\bar n_1$ and $\bar n_2$. It can be seen that, as the average phonon number increases from 0 to 0.2, the infidelity increases from $4.4\times 10^{-4}$ to $1.7 \times 10^{-3}$  at an approximately linear rate. In summary,  the optimal pulse is smooth, robust, and has   amplitudes not exceeding 7 MHz, thus can be friendly to experimental implementation.

\section{Conclusion and Outlook}
QOC is   powerful     for designing  shaped pulses to induce target quantum operations. The idea of employing  QOC   for realizing  fast entangling   gates  on trapped-ions   system is   simple and direct,   but was not yet explored. The reason might be that the laser-ion interaction Hamiltonian in the fast gate regime has strong nonlinearity, which looks not   easy to deal with. Our developed framework here    allows for conveniently pursue high-quality shaped optical fields to drive the ions in the nonlinear regime. It remarkably does not depend on the assumption of   Lamb-Dicke approximation, nor the application of average Hamiltonian theory. Through numerical tests on  the case of two-qubit  fast gate,   we   learned that: in ideal conditions, accurate and fast  gates within one period of the trap frequency do exist, which suggests   the speed limit to quantum processing with ions; but to meet current experimental requirements, longer pulses are necessary. Our results of $1$ $\mu$s- and $3$ $\mu$s-gates indicate that  as the gate time increases, higher-fidelity and lower-intensity optimal pulse can be found. As a further example, if we   increase the gate time to  $5$ $\mu$s, we can achieve phase-insensitive gates of fidelity above  0.9999 with pulse amplitudes below 5 MHz.
Therefore, it is to the experimenters' choice to decide   which of the following, gate speed, fidelity, phase robustness and other factors, are the most important considerations in real experiments.

The generic method introduced here may potentially advance small-sized trapped ion  systems toward a new level of gate speed as well as precision. As for larger-sized   systems, however, there is   one important yet challenging problem. That is, QOC suffers from the intrinsic scalability issue. As the number of ion qubits grows, the system dimension grows exponentially, so it will require huge amount of computational resources to simulate the full controlled dynamics. Therefore,  future work should place greater focus on improving the    simulation efficiency. To give an example,  we mention a software library  introduced in Ref. \cite{Ilya11}, that trying to construct  accurate low-dimensional matrix representations to replace the full evolution operators,     and hence allows       pulse search on large spin systems of as many as 40 spins  with relative ease    just on a desktop workstation. Our framework could incorporate similar strategies so as to extend  to,  e.g.,  long ion chains.  Finally, the present framework is not limited to   quantum gate implementation, but can also be applied to other quantum information processing tasks, which is   worth further investigation.  We anticipate that the systematic method proposed in this work can find experimental implementations as well as more quantum technology applications.

\section{Acknowledgments}
We acknowledge support by   the National Natural Science Foundation of China (Grants No. 1212200199,   11975117, 12004165, 92065111, and 12204230), Guangdong Basic and Applied Basic Research Foundation (Grant No. 2021B1515020070),  Guangdong Provincial Key Laboratory (Grant No. 2019B121203002), and Shenzhen Science and Technology Program (Grants No. RCYX20200714114522109, RCBS20200714114820298,  and     KQTD20200820113010023). Y.L. was supported by the National Natural Science Foundation of China (Grants No. 92165206, No. 11974330),  Innovation Program for Quantum Science and Technology (Grant No. 2021ZD0301603), and the Fundamental Research Funds for the Central Universities.


%

\end{document}